\documentclass[11pt]{article}
\usepackage{amsfonts}
 \usepackage{pifont}
\usepackage{amsmath,amsfonts,amssymb,theorem,graphicx,listings,txfonts}
\usepackage{graphics}
 \usepackage{caption2}
\usepackage{flafter}
 \usepackage{indentfirst}
\usepackage{epsfig}
  \usepackage{mathrsfs}
\usepackage{subfigure}
 \usepackage{cite}
 \usepackage{float}
\newtheorem{theorem}{Theorem}[section]
\newtheorem{lemma}[theorem]{Lemma}

\newtheorem{proposition}[theorem]{Proposition}

\newtheorem{definition}[theorem]{Definition}

\theoremstyle{example}
\newtheorem{example}[theorem]{Example}
\theoremstyle{procedure}

\theoremstyle{problem}

\title{ Some new similarity measures for hesitant fuzzy sets and their applications in multiple attribute decision making \thanks{This work is supported by the National
Natural Science Foundation of China (NO. 11071061, 11101135) and the
National Basic Research Program of China(NO. 2011CB311808).}}

\author{\small Xiaoqiang Zhou$^{a,b}$, Qingguo Li$^{a}$\thanks{Corresponding author. \quad Tel./fax: +86 0731 88822755.
\newline\mbox{}\hspace{0.55cm}
E-mail address: zxq0923@163.com, liqingguoli@yahoo.com.cn.}~\\
{\small $^{a}$College of Mathematics and Econometrics, Hunan University}\\
{\small Changsha, 410082, P.R.China}\\
{\small $^{b}$College of Mathematics,Hunan Institute of Science and
Technology}\\
{\small Yueyang, 414006, P.R.China}\\
}
\date{}

\begin{document}
\maketitle \baselineskip=17pt
\begin{center}
\begin{quote}
 {{\bf Abstract.}
Similarity measure is a very important topic in fuzzy set theory. Torra (2010) proposed the notion of hesitant fuzzy set(HFS), which is a generalization of the notion of Zadeh' fuzzy set. In this paper, some new similarity measures for HFSs are developed. Based on the proposed similarity measures, a method of multiple attribute decision making under hesitant fuzzy environment is also introduced. Additionally, a numerical example is given to illustrate the application of the proposed similarity measures of HFSs to decision-making.
\\}
\end{quote}
\end{center}
\renewcommand{\thesection}{\arabic{section}}
\section{Introduction}
Ever since the notion of fuzzy set was given by Zadeh \cite{zadeh1}, many new theories and approaches dealing with uncertainty and imprecision have been introduced. Some of them, such as intuitionistic fuzzy set(IFS) \cite{atanassov1}, interval-valued fuzzy set(IVFS) \cite{atanassov2}, vague set \cite{gau1}, and type-2 fuzzy set(T2FS) \cite{zadeh2}, are extensions of ordinary fuzzy set theory. After that, many researchers have studied this topic and obtained a lot of meaningful results in cluster analysis, multi-criteria decision, aggregation and grey relational analysis.
\par
As is well known, the similarity measure is a very important concept, for it provides the degree of similarity between two fuzzy objects. Since Zadeh \cite{zadeh3} introduced the similarity
relation concept, similarity measures of fuzzy sets have been widely studied from different aspects and applied in various fields, such as decision-making, cluster analysis, machine learning, market prediction, approximate reasoning, image processing, and pattern recognition. Fan and Xie \cite{fan1} as well as Liu\cite{liu2} gave the axiom definition and studied some properties of similarity measures between fuzzy sets. Pappis and Karacapilidis \cite{pappis1} investigated three similarity measures of fuzzy sets based on intersection and union operations, the maximum difference and the differences as well as the sum
of membership grades. In \cite{wang1}, Wang proposed two new similarity measures between fuzzy sets and between elements. Turksen and Zhong \cite{turksen1} applied similarity measures
of fuzzy sets for an approximate analogical reasoning. In a
multimedia database query, Candan et al. \cite{candan1} applied similarity measures to develop
query processing with different fuzzy semantics. Moreover, lots of similarity measures for IFSs, IVFSs, vague sets and T2FSs have also been widely developed in the literatures \cite{hung1,hung2,hung3,li1,li2,liang1,liu1,xu3,gau1,chen3,chen4,hung4,yang2,zhang1,zeng1}.
\par
Recently, to deal with hesitant and incongruous problems, Torra and Narukawa \cite{torra1,torra2} introduced the concept of hesitant fuzzy set(HFS), which is also an extension of the classic fuzzy set, for it permits the membership
degree of an element to a set to be represented as several possible values between 0 and 1. After the pioneering work of Torra, the HFS has received much attention from many authors and has been used in decision-making and clustering analysis \cite{chen2,rodriguez1,wei1,xia1,xia2,xu1,xu2,xu5,yu1,zhu1}. For example, Chen \cite{chen2} systematically investigated the correlation coefficients of HFSs and applied them to clustering analysis, Xia and Xu \cite{xia1} studied the aggregation
operators of hesitant fuzzy sets and applied them to decision making. Xu and Xia \cite{xu1} gave the axiom definitions of distance and similarity measures between HFSs. they also presented some distance measures for HFSs and obtained some similarity measures corresponding to the distances of HFSs. However, their axiom definitions of distance and similarity measures only satisfy three properties, respectively. The more reasonable definitions of distance and similarity measures, in general, should have four properties like the notions of fuzzy sets \cite{fan1,liu2}, IFSs \cite{hung3,li1,mitchell1}, IVFSs \cite{zeng1} and T2FSs \cite{hung4}.  Therefore, in this paper we modify the axiom definitions of distance and similarity measures for HFSs and propose some new distance and similarity measures between HFSs.
\par
The rest of this paper is organized as follows. In Section 2, we review the notions of HFS and give the modified axiom definitions of distance and similarity measures for HFSs. In Section 3, we present some new geometric distance and similarity measures between HFSs base on geometric distance model and set-theory approach. We apply the proposed similarity measures of HFSs to hesitant fuzzy decision-making in Section 4. We make the conclusions in Section 5.

\section{Preliminaries}
 In this section, we briefly recall the necessary definitions and notations of HFS and modify the axiom definitions of distance and similarity measures between HFSs, which were first given by Xu and Xia \cite{xu1}.\par
$HFSs$ are very useful in
dealing with the situations where people have hesitation in providing their preferences over objects in a decision-making
process. The definition of HFS was first introduced by Torra and Narukawa \cite{torra1,torra2} as follows.
\begin{definition}
Let X be a reference set, an HFS on X is in terms of a function that when applied to X returns a subset of $[0,1]$, which can be represented as
$H=\{ \frac{h_{H}(x)}{x}|x\in X\}$,
where $h_{H}(x)$ is a set of some values in $[0,1]$, denoting the possible membership degrees of the element $x\in X$ to the set H.
\end{definition}
For convenience, Xu and Xia \cite{xu1} called $h_{H}(x)$ an hesitant fuzzy element(HFE) with respect to $x$ of $H$. It is worth noting that the number of values of different HFEs may be different, let $n(h_{H}(x))$ be the number of values of $h_{H}(x)$. We arrange the values of $h_{H}(x)$ in decreasing order, and
let $h_{H}^{\sigma(i)}(x)$ be the $i$th smallest value of $h_{H}(x)$.\par
Distance and similarity measures are the fundamental and important issues of theory of sets. For HFSs, the axiom definitions of distance and similarity measures were first addressed by Xu and Xia \cite{xu1}.
\begin{definition}\label{def1}
Let $A$ and $B$ be two $HFSs$ on $X=\{x_{1},x_{2},\cdots ,x_{m}\}$. Then the distance measure between $A$ and $B$ is defined as $d(A,B)$,
which satisfies the following properties:\\
$(1)$ $0\leq d(A,B)\leq 1$;\\
$(2)$ $d(A,B)=0\Leftrightarrow A=B$;\\
$(3)$ $d(A,B)=d(B,A)$.
\end{definition}
\begin{definition}\label{def2}
Let $A$ and $B$ be two $HFSs$ on $X=\{x_{1},x_{2},\cdots ,x_{m}\}$. Then the similarity measure between $A$ and $B$ is defined as $s(A,B)$,
which satisfies the following properties:\\
$(1)$ $0\leq s(A,B)\leq 1$;\\
$(2)$ $s(A,B)=1\Leftrightarrow A=B$;\\
$(3)$ $s(A,B)=s(B,A)$.
\end{definition}
In many cases, however, $n(h_{A}(x))\neq n(h_{B}(x))$. To operate correctly, it is requested that two HFEs have the same length when they are compared. Thus we should extend the shorter one such that their length is the same. For this,  Xu and Xia \cite{xu1} give the following regulation:
\par If $n(h_{A}(x))>n(h_{B}(x))$, then $h_{B}(x)$ should be extended by adding the minimum value in it until it has the same length with $h_{A}(x)$; If $n(h_{A}(x))<n(h_{B}(x))$, then $h_{A}(x)$ should be extended by adding the minimum value in it until it has the same length with $h_{B}(x)$. For instance, let $h_{A}(x)=\{0.5,0.4\}$, $h_{B}(x)=\{0.7,0.4,0.2\}$. Clearly, $n(h_{A}(x))<n(h_{2}(x))$, so we should
extend $h_{A}(x)$ to $h_{A}(x)=\{0.5,0.4,0.4\}$.
\par In fact, we can extend the shorter
HFE by adding any value in it until it has the same length with the longer one according to the decision makers' preferences and
actual situations. In this paper, we assume
that the decision makers all adopt the above regulation.
\par Based on the above regulation, we define the following comparison laws.
\begin{definition}\label{def00}
Let $A$ and $B$ be two $HFSs$ on $X$, and $n_{x}=max\{n(h_{A}(x)), n(h_{B}(x))\}$ for all $x\in X$. Then \\
$(1)$ $h_{A}(x)$ is said to be inferior to $h_{B}(x)$, denoted by $h_{A}(x)\preceq h_{B}(x)$, if $h_{A}^{\sigma(i)}(x))\leq h_{B}^{\sigma(i)}(x)$ for all $i=1,2,\cdots,n_{x}$. Especially, if $n_{x}=n(h_{A}(x))=n(h_{B}(x))$ and $h_{A}^{\sigma(i)}(x))\leq h_{B}^{\sigma(i)}(x)$ for all $i=1,2,\cdots,n_{x}$, then $h_{A}(x)$ is said to be less than $h_{B}(x)$, denoted by $h_{A}(x)\leq h_{B}(x)$. \\
$(2)$ $h_{A}(x)$ is said to be equal to $h_{B}(x)$ if $h_{A}^{\sigma(i)}(x))=h_{B}^{\sigma(i)}(x)$ for all $i=1,2,\cdots,n_{x}$, denoted by $h_{A}(x)=h_{B}(x)$.\\
$(3)$ $HFS$ $A$ is said to be an hesitant fuzzy quasi subset of $HFS$ $B$, denoted by $A\sqsubseteq B$, if $h_{A}(x)\preceq h_{B}(x)$ for all $x\in X$. Especially, if $h_{A}(x)\leq h_{B}(x)$ for all $x\in X$, then $A$ is called an hesitant fuzzy subset of $B$, denoted by $A\subseteq B$.\\
$(4)$ $HFS$ $A$ is said to be equal to  $HFS$ $B$, denoted by $A=B$, if $h_{A}(x)=h_{B}(x)$ for all $x\in X$.
\end{definition}
\begin{proposition}
  Let $A$ and $B$ be two $HFSs$ on $X$. If $h_{A}(x)=h_{B}(x)$, then $n(h_{A}(x))=n(h_{B}(x))$.
\end{proposition}
\noindent\textbf{Proof.}
  The proof is easily obtained from Definition \ref{def00}.$\Box$\par
Based on Definition \ref{def00}, we modify the axiom definitions of the distance and similarity measures as follows:
\begin{definition}\label{def11}
Let $A$ and $B$ be two $HFSs$ on $X=\{x_{1},x_{2},\cdots ,x_{m}\}$, and $d_{max}=max\{d(A,B)\}$. Then the distance measure between $A$ and $B$ is defined as $d(A,B)$,
which satisfies the following properties:\\
$(D1)$ $0\leq d(A,B)\leq d_{max}$;\\
$(D2)$ $d(A,B)=0\Leftrightarrow A=B$;\\
$(D3)$ $d(A,B)=d(B,A)$;\\
$(D4)$ Let $C$ be an HFS, if $A\sqsubseteq B\sqsubseteq C$, then $d(A,B)\leq d(A,C)$ and $d(B,C)\leq d(A,C)$.
\end{definition}
If $(D1')$ replaces $(D1)$, then $d(A,B)$ is called a normalized distance measure, where\\
$(D1')$ $0\leq d(A,B)\leq 1$.
\begin{definition}\label{def12}
Let $A$ and $B$ be two $HFSs$ on $X=\{x_{1},x_{2},\cdots ,x_{m}\}$. Then the similarity measure between $A$ and $B$ is defined as $s(A,B)$,
which satisfies the following properties:\\
$(P1)$ $0\leq s(A,B)\leq 1$;\\
$(P2)$ $s(A,B)=1\Leftrightarrow A=B$;\\
$(P3)$ $s(A,B)=s(B,A)$;\\
$(P4)$ Let $C$ be an HFS, if $A\sqsubseteq B\sqsubseteq C$, then $s(A,C)\leq (A,B)$ and $s(A,C)\leq s(B,C)$.
\end{definition}
\section{Some new similarity measures for hesitant fuzzy sets}
 Let $A$ and $B$ be two $HFSs$ on $X=\{x_{1},x_{2},\cdots ,x_{m}\}$. In this section, we introduce some new distance and similarity measures between hesitant fuzzy sets.
\subsection{Similarity measures based on geometric distance model}
 Xu and Xia\cite{xu1} introduced a lot of geometric distance models between hesitant fuzzy sets $A$ and $B$. Some of them are given as follows:\\
$(1)$ Hesitant normalized Hamming distance:
\begin{eqnarray}\label{eqnd1}
d_{1}(A,B)=\frac{1}{m}\sum \limits_{i=1}^m{\left(\frac{1}{n_{x_{i}}}\sum\limits_{j=1}^{n_{x_{i}}}
{|h_{A}^{\sigma(j)}(x_{i})-h_{B}^{\sigma(j)}(x_{i})|}\right)}
\end{eqnarray}
$(2)$ Hesitant normalized Euclidean distance:
\begin{eqnarray}\label{eqnd2}
d_{2}(A,B)=\sqrt{\frac{1}{m}\sum \limits_{i=1}^m{\left(\frac{1}{n_{x_{i}}}\sum\limits_{j=1}^{n_{x_{i}}}
{|h_{A}^{\sigma(j)}(x_{i})-h_{B}^{\sigma(j)}(x_{i})|^{2}}\right)}}
\end{eqnarray}
$(3)$ Generalized hesitant normalized distance:
\begin{eqnarray}\label{eqnd3}
d_{3}(A,B)=\left[\frac{1}{m}\sum \limits_{i=1}^m{\left(\frac{1}{n_{x_{i}}}\sum\limits_{j=1}^{n_{x_{i}}}
{|h_{A}^{\sigma(j)}(x_{i})-h_{B}^{\sigma(j)}(x_{i})|^{p}}\right)}\right]^{1/p}, p>0.
\end{eqnarray}
 Clearly, If $p=1$, then Eq. (\ref{eqnd3}) is reduced to Eq. (\ref{eqnd1}).
\par From Eq. (\ref{eqnd1}), we know that
\begin{eqnarray*}
d_{i}=\frac{1}{n_{x_{i}}}\sum\limits_{j=1}^{n_{x_{i}}}
{|h_{A}^{\sigma(j)}(x_{i})-h_{B}^{\sigma(j)}(x_{i})|}
\end{eqnarray*}
indicates the distance between the $i$th HFE of $A$ and $B$, and $d_{1}(A,B)$ indicates the mean of distances between all elements of $A$ and $B$. From the point of view, we define another generalized normalized distance of $A$ and $B$ as:
\begin{eqnarray}\label{eqnd4}
d_{4}(A,B)=\frac{1}{m}\sum \limits_{i=1}^m{\left(\frac{1}{n_{x_{i}}}\sum\limits_{j=1}^{n_{x_{i}}}
{|h_{A}^{\sigma(j)}(x_{i})-h_{B}^{\sigma(j)}(x_{i})|^{p}}\right)}^{1/p}, p>0,
\end{eqnarray}
which we call type-2 generalized hesitant normalized distance.
It is clear that Eq. (\ref{eqnd4}) is different from Eq. (\ref{eqnd3}). But if $p=1$, then Eq. (\ref{eqnd4}) is also reduced to Eq. (\ref{eqnd1}). If $p=2$, then Eq. (\ref{eqnd4}) becomes type-2 hesitant normalized Euclidean distance:
\begin{eqnarray}\label{eqnd84}
d_{5}(A,B)=\frac{1}{m}\sum \limits_{i=1}^m{\sqrt{\frac{1}{n_{x_{i}}}\sum\limits_{j=1}^{n_{x_{i}}}
{|h_{A}^{\sigma(j)}(x_{i})-h_{B}^{\sigma(j)}(x_{i})|^{2}}}}.
\end{eqnarray}
Then it is natural to ask ``Is the defined distance $d_{4}(A,B)$
reasonable?". We answer this question in Theorem \ref{theo1}.
\begin{theorem}\label{theo1}
$d_{4}(A,B)$ is a normalized distance measure between HFSs $A$ and $B$.
\end{theorem}
\noindent\textbf{Proof.}
It is easy to see that $d_{4}(A,B)$ satisfies the properties $(D1')-(D3)$. We therefore only prove $(D4)$. Let $A\sqsubseteq B\sqsubseteq C$, then $h_{A}(x_{i})\preceq h_{B}(x_{i})\preceq h_{C}(x_{i})$ for each $x_{i}\in X$. It follows that
\begin{eqnarray*}
\begin{split}
&|h_{A}^{\sigma(j)}(x_{i})-h_{B}^{\sigma(j)}(x_{i})|^{p}\leq |h_{A}^{\sigma(j)}(x_{i})-h_{C}^{\sigma(j)}(x_{i})|^{p},\\ &|h_{B}^{\sigma(j)}(x_{i})-h_{C}^{\sigma(j)}(x_{i})|^{p}\leq |h_{A}^{\sigma(j)}(x_{i})-h_{C}^{\sigma(j)}(x_{i})|^{p},\\
\Rightarrow &\frac{1}{n_{x_{i}}}\sum\limits_{j=1}^{n_{x_{i}}}{|h_{A}^{\sigma(j)}(x_{i})-h_{B}^{\sigma(j)}(x_{i})|^{p}}\leq \frac{1}{n_{x_{i}}}\sum\limits_{j=1}^{n_{x_{i}}}{|h_{A}^{\sigma(j)}(x_{i})-h_{C}^{\sigma(j)}(x_{i})|^{p}},\\ &\frac{1}{n_{x_{i}}}\sum\limits_{j=1}^{n_{x_{i}}}{|h_{B}^{\sigma(j)}(x_{i})-h_{C}^{\sigma(j)}(x_{i})|^{p}}\leq \frac{1}{n_{x_{i}}}\sum\limits_{j=1}^{n_{x_{i}}}{|h_{A}^{\sigma(j)}(x_{i})-h_{C}^{\sigma(j)}(x_{i})|^{p}},\\
\Rightarrow &d_{4}(A,B)\leq d_{4}(A,C), d_{4}(B,C)\leq d_{4}(A,C).
\end{split}
\end{eqnarray*}
Thus the property $(D4)$ is obtained.
$\Box$\par
Based on Eq. (\ref{eqnd4}), we further define type-2 generalized hesitant distances as follows:
\begin{eqnarray}\label{eqnd54}
d_{5}(A,B)=\sum \limits_{i=1}^m{\left(\frac{1}{n_{x_{i}}}\sum\limits_{j=1}^{n_{x_{i}}}
{|h_{A}^{\sigma(j)}(x_{i})-h_{B}^{\sigma(j)}(x_{i})|^{p}}\right)}^{1/p}, p>0.
\end{eqnarray}
\begin{eqnarray}\label{eqnd55}
d_{6}(A,B)=\frac{1}{m}\sum \limits_{i=1}^m{\left(\sum\limits_{j=1}^{n_{x_{i}}}
{|h_{A}^{\sigma(j)}(x_{i})-h_{B}^{\sigma(j)}(x_{i})|^{p}}\right)}^{1/p}, p>0.
\end{eqnarray}
\begin{eqnarray}\label{eqnd56}
d_{7}(A,B)=\sum \limits_{i=1}^m{\left(\sum\limits_{j=1}^{n_{x_{i}}}
{|h_{A}^{\sigma(j)}(x_{i})-h_{B}^{\sigma(j)}(x_{i})|^{p}}\right)}^{1/p}, p>0.
\end{eqnarray}
\begin{theorem}\label{theo21}
$d_{i}(A,B)(i=5,6,7)$ is a distance measure between HFSs $A$ and $B$, and satisfies the following properties:\\
$(1)$ $0\leq d_{5}(A,B)\leq m$;\\
$(2)$ $0\leq d_{6}(A,B)\leq \frac{1}{m}\sum\limits_{i=1}^{m}{(n_{x_{i}})^{1/p}}$;\\
$(3)$ $0\leq d_{7}(A,B)\leq \sum\limits_{i=1}^{m}{(n_{x_{i}})^{1/p}}$.
\end{theorem}
\noindent\textbf{Proof.}
The proof of $(D2)-(D4)$ is  similar to Theorem \ref{theo1}, We only prove $(1)-(3)$. Let $h_{A}^{\sigma(j)}(x_{i})=1$ and $h_{B}^{\sigma(j)}(x_{i})=0$ for all $x_{i}\in X$ and $j=1,2,\cdots,n_{x_{i}}$, then $d_{5}(A,B)=m$, $d_{6}(A,B)=\frac{1}{m}\sum\limits_{i=1}^{m}{(n_{x_{i}})^{1/p}}$ and $d_{7}(A,B)=\sum\limits_{i=1}^{m}{(n_{x_{i}})^{1/p}}$.
$\Box$\par
The $L_{P}$ metric is very important and has been used to measure the distance of fuzzy sets and $IFSs$ \cite{hung1}. If we apply the $L_{P}$ metric to the distance measure between $HFSs$, then a hesitant $L_{P}$ distance is
given as
\begin{eqnarray}\label{eqnd6}
  d_{8}(A,B)=\frac{1}{m}\sum \limits_{i=1}^m{\left(\sum\limits_{j=1}^{n_{x_{i}}}
{|h_{A}^{\sigma(j)}(x_{i})-h_{B}^{\sigma(j)}(x_{i})|^{p}}\right)}^{1/p}, p\geq 1.
\end{eqnarray}
Clearly, if $p\geq 1$, then the type-2 generalized hesitant distance $d_{6}(A,B)$ becomes the hesitant $L_{p}$ distance $d_{8}(A,B)$.
\par However, there is an interesting result: if $p\rightarrow \infty$, then the hesitant $L_{p}$ distance $d_{8}(A,B)$ is reduced to hesitant normalized Hamming-Hausdorff distance
\begin{eqnarray}
d_{9}(A,B)=\frac{1}{m}\sum \limits_{i=1}^m{\max\limits_{j}
|h_{A}^{\sigma(j)}(x_{i})-h_{B}^{\sigma(j)}(x_{i})|},
\end{eqnarray}
which is defined by Xu and Xia \cite{xu1}.\par
To prove the above result, the following lemma is needed.
\begin{lemma}\label{lemma3}
  Let $a_{i}\in \mathbb{R}$ and $a_{i}\geq 0,i=1,2,\cdots,k$. Then
\begin{eqnarray*}
  \lim\limits_{p\to \infty}{(a_{1}^p+a_{2}^p+\cdots +a_{k}^p)^{1/p}}=\max\limits_{i}\{a_{i}\}, p\geq 1.
\end{eqnarray*}
\end{lemma}
\noindent\textbf{Proof.}
  It is obvious whenever (i) $a_{i}=0(i=1,2,\cdots,k)$, or (ii) $a_{1}=a_{2}=\cdots=a_{k}$, because $\lim\limits_{p\to \infty}{k^{1/p}}=1$. If $a_{i}\neq a_{j},i\neq j, i,j=1,2,\cdots,k$, then the following show that
  \begin{eqnarray*}
  \lim\limits_{p\to \infty}{(a_{1}^p+a_{2}^p+\cdots +a_{k}^p)^{1/p}}=\max\limits_{i}\{a_{i}\}.
\end{eqnarray*}
  Without loss of generality, we suppose that $a_{1}\geq a_{2}\geq \cdots \geq a_{k}$, and let $y=(a_{1}^p+a_{2}^p+\cdots +a_{k}^p)^{1/p}$. Then
  \begin{eqnarray*}
  \lim\limits_{p\to \infty}lny=
  \lim\limits_{p\to \infty}\frac{a_{1}^p+a_{2}^p +\cdots +a_{k}^p}{p}.
\end{eqnarray*}
   Using L'Hospital's rule, we have
\begin{eqnarray*}
\begin{split}
  \lim\limits_{p\to \infty}lny &=
  \lim\limits_{p\to \infty}\frac{a_{1}^p lna_{1}+a_{2}^p lna_{2}+\cdots +a_{k}^p lna_{k}}{a_{1}^p+a_{2}^p+\cdots +a_{k}^p}\\
  &=\lim\limits_{p\to \infty}\frac{(lna_{1}+(a_{2}/a_{1})^p lna_{2}+\cdots +(a_{k}/a_{1})^p lna_{k}}{1+(a_{2}/a_{1})^p+\cdots +(a_{k}/a_{1})^p}\\
  &=lna_{1}.
\end{split}
\end{eqnarray*}
Therefore,
\begin{eqnarray*}
\lim\limits_{p\to \infty}y=\lim\limits_{p\to \infty}{(a_{1}^p+a_{2}^p+\cdots +a_{k}^p)^{1/p}}=a_{1}=\max\limits_{i}\{a_{i}\}.\Box
\end{eqnarray*}
\begin{theorem}$
  \lim\limits_{p \to \infty}{d_{8}(A,B)}=\frac{1}{m}\sum \limits_{i=1}^m{\max\limits_{j}
|h_{A}^{\sigma(j)}(x_{i})-h_{B}^{\sigma(j)}(x_{i})|}$.
\end{theorem}
\noindent\textbf{Proof.}
  It can be obtained directly by Lemma \ref{lemma3}.
$\Box$
\par In many practical problems, however, the weight of the element $x_{i}\in X$ should be taken into account. Especially for multiple attribute decision making problems, the considered attributes usually are of different importance. Thus we need to consider the weight of the element so that
we get the following weighted distance between HFSs. Assume that $w_{i}(i=1,2,\cdots,m)$ is the weight of the element $x_{i}\in X$,$w_{i}\in [0,1]$ and $\sum\limits_{i=1}^{m}{w_{i}}=1$, then we obtain a type2-generalized hesitant weighted distance
\begin{eqnarray}\label{eqnd8}
  d_{10}(A,B)=\sum \limits_{i=1}^m{w_{i}\left(\frac{1}{n_{x_{i}}}\sum\limits_{j=1}^{n_{x_{i}}}
{|h_{A}^{\sigma(j)}(x_{i})-h_{B}^{\sigma(j)}(x_{i})|^{p}}\right)}^{1/p}, p>0.
\end{eqnarray}
and a hesitant $L_{p}$ weighted disatance
\begin{eqnarray}\label{eqnd9}
 d_{11}(A,B)=\sum \limits_{i=1}^m{w_{i}\left(\sum\limits_{j=1}^{n_{x_{i}}}
{|h_{A}^{\sigma(j)}(x_{i})-h_{B}^{\sigma(j)}(x_{i})|^{p}}\right)}^{1/p}, p\geq 1.
\end{eqnarray}
Obviously, if each element has the same importance, that is, $w_{i}=1/m, (i=1,2,\cdots,n)$, then the Eq.s (\ref{eqnd8}) and (\ref{eqnd9}) are reduced to Eq.s (\ref{eqnd4}) and (\ref{eqnd6}), respectively.\par
It is seen that all the above distance measures are discrete, if both the universe of discourse and the weight of element are continuous, then we get the continuous distances. Let the weight of $x\in X=[a,b]$ be $w(x)$ with $w(x)\in [0,1]$ and $\int_{a}^{b}{w(x)dx}=1$, we define a type-2 continuous hesitant weighted Euclidean distance and type-2 generalized continuous hesitant weighted distance as follows, respectively:
\begin{eqnarray}
  d_{12}(A,B)=\int_{a}^{b}{w(x)\left[\frac{1}{n_{x}}\sum\limits_{j=1}^{n_{x}}
{|h_{A}^{\sigma(j)}(x)-h_{B}^{\sigma(j)}(x)|^{2}}\right]^{1/2}dx}
\end{eqnarray}
\begin{eqnarray}
  d_{13}(A,B)=\int_{a}^{b}{w(x)\left[\frac{1}{n_{x}}\sum\limits_{j=1}^{n_{x}}
{|h_{A}^{\sigma(j)}(x)-h_{B}^{\sigma(j)}(x)|^{p}}\right]^{1/p}dx},p>0.
\end{eqnarray}
Especially, if $w(x)=1/(b-a)$ for all $x\in [a,b]$, then the type-2 continuous hesitant weighted Euclidean distance is reduced to a type-2 continuous hesitant normalized Euclidean distance
\begin{eqnarray}
  d_{14}(A,B)=\frac{1}{(b-a)}\int_{a}^{b}{\left[\frac{1}{n_{x}}\sum\limits_{j=1}^{n_{x}}
{|h_{A}^{\sigma(j)}(x)-h_{B}^{\sigma(j)}(x)|^{2}}\right]^{1/2}dx}
\end{eqnarray}
and the type-2 generalized continuous hesitant weighted distance is reduced to a type-2 generalized continuous hesitant normalized distance
\begin{eqnarray}
  d_{15}(A,B)=\frac{1}{(b-a)}\int_{a}^{b}{\left[\frac{1}{n_{x}}\sum\limits_{j=1}^{n_{x}}
{|h_{A}^{\sigma(j)}(x_{i})-h_{B}^{\sigma(j)}(x)|^{p}}\right]^{1/p}dx},p>0.
\end{eqnarray}
Based on $L_{p}$ metric, we define a continuous hesitant weighted $L_{p}$ distance
\begin{eqnarray}
  d_{16}(A,B)=\int_{a}^{b}{w(x)\left[\sum\limits_{j=1}^{n_{x}}
{|h_{A}^{\sigma(j)}(x)-h_{B}^{\sigma(j)}(x)|^{p}}\right]^{1/p}dx},p\geq 1.
\end{eqnarray}
Especially, if $w(x)=1/(b-a)$ for all $x\in [a,b]$, then the continuous hesitant weighted $L_{p}$ distance is reduced to a continuous hesitant average $L_{p}$ distance
\begin{eqnarray}
  d_{17}(A,B)=\frac{1}{(b-a)}\int_{a}^{b}{\left[\sum\limits_{j=1}^{n_{x}}
{|h_{A}^{\sigma(j)}(x)-h_{B}^{\sigma(j)}(x)|^{p}}\right]^{1/p}dx},p\geq 1.
\end{eqnarray}
Motivated by the ordered weighted idea \cite{xu4}, similar to literature \cite{xu1}, we can get the hesitant ordered weighted distances corresponding to aforementioned distances.
\par
As is well known, an exponential operation is
very useful in dealing with the similarity relation \cite{zadeh3}, classical Shannon entropy \cite{pal1} and in cluster analysis \cite{yang1}.
We therefore adopted the exponential operation to a distance of HFSs and get a new distance measure between HFSs. Let $d(A,B)$ be a distance between HFSs $A$ and $B$ and $d_{max}=max\{d(A,B)\}$, then we define an exponential-type distance measure:
\begin{eqnarray}\label{eqnd99}
d_{18}(A,B)=\frac{1-exp(-d(A,B))}{1-exp(-d_{max})}
\end{eqnarray}
we give the following lemma to prove Eq. (\ref{eqnd99}) is a reasonable distance measure.
\begin{lemma}\label{lemma8}
Let $f(x)=\frac{1-exp(-x)}{1-exp(-m)}, x\in [0,m]$, then $f_{min}(x)=f(0)=0$ and $f_{max}(x)=f(m)=1$.
\end{lemma}
\noindent\textbf{Proof.}
  Since $f^{'}(x)=\frac{exp(-x)}{1-exp(-m)}> 0, x\in [0,m]$, then $f(x)$ is increasing in $[0,m]$.
$\Box$
\begin{theorem} \label{theorem10}
Let $d(A,B)$ be a distance between HFSs $A$ and $B$, and $d_{max}=max\{d(A,B)\}$.
Then $d_{18}(A,B)$ is a normalized distance measure of HFSs $A$ and $B$.
\end{theorem}
\noindent\textbf{Proof.} $(D1')-(D3)$ is easily obtained, We only prove $(D4)$. Since $d(A,B)$ is a distance measure between HFSs $A$ and $B$, then $d(A,B)\leq d(A,C)$ and $d(B,C)\leq d(A,C)$ for $A\sqsubseteq B\sqsubseteq C$. By Lemma \ref{lemma8}, we have $d_{18}(A,B)\leq d_{18}(A,C)$ and $d_{18}(B,C)\leq d_{18}(A,C)$ for $A\sqsubseteq B\sqsubseteq C$. Thus the property (D4) is obtained.
$\Box$\par
  From Theorem \ref{theorem10}, we know that $d_{18}(A,B)$ is a normalized distance of $d(A,B)$, that is to say, we can use Eq. (\ref{eqnd99}) to generate a normalized distance of $d(A,B)$.
\par It is will known that the similarity measure and distance measure are dual concepts. Hence we may use a distance measure to define a similarity measure.
\begin{theorem}\label{theo10}
Let $A$ and $B$ be HFSs. Let $f$ be a monotone decreasing function, $d$ a distance measure and $d_{max}$ the maximal distance. We define
\begin{eqnarray}\label{eqna200}
  s_{0}(A,B)=\frac{f(d(A,B))-f(d_{max})}{f(0)-f(d_{max})},
\end{eqnarray}
  then $s_{0}(A,B)$ is a similarity measure between HFSs $A$ and $B$.
\end{theorem}
\noindent\textbf{Proof.}
$(1)$ Since $f$ be a monotone decreasing function and $0\leq d(A,B)\leq d_{max}$, then $f(d_{max})\leq f(d(A,B))\leq f(0)$. This implies
\begin{eqnarray*}
  0\leq \frac{f(d(A,B))-f(d_{max})}{f(0)-f(d_{max})}\leq 1.
\end{eqnarray*}
$(2)$  $d(A,B)=0\Leftrightarrow A=B$ implies $s_{0}(A,B)=1\Leftrightarrow A=B$.\\
$(3)$ $d(A,B)=d(B,A)$ implies $s_{0}(A,B)=s_{0}(B,A)$.\\
$(4)$ Let $C$ be an HFS, and $A\sqsubseteq B\sqsubseteq C$, then $d(A,B)\leq d(A,C)$ and $d(B,C)\leq d(A,C)$. Since $f$ be a monotone decreasing function, then $f(d(A,C))\leq f(d(A,B))$ and $f(d(A,C))\leq f(d(B,C))$. These imply $s_{0}(A,C)\leq s_{0}(A,B)$ and $s_{0}(A,C)\leq s_{0}(B,C)$.
$\Box$\par
By Theorem \ref{theo10}, if we choose $f(x)=1-x($or $e^{-x}$ or $\frac{1}{1+x})$, then the corresponding similarity measures between $A$
and $B$ can be obtained. For example, let $f(x)=1-x$, then $s_{0}(A,B)=1-\frac{d(A,B)}{d_{max}}$. Based on Eq.s (\ref{eqnd3}), (\ref{eqnd4}) and (\ref{eqnd55}), we obtain the similarity measures corresponding to the distance measures as follows, respectively:
\begin{eqnarray}\label{eqns3}
s_{1}(A,B)=1-\left[\frac{1}{m}\sum \limits_{i=1}^m{\left(\frac{1}{n_{x_{i}}}\sum\limits_{j=1}^{n_{x_{i}}}
{|h_{A}^{\sigma(j)}(x_{i})-h_{B}^{\sigma(j)}(x_{i})|^{p}}\right)}\right]^{1/p},
\end{eqnarray}
\begin{eqnarray}\label{eqns4}
s_{2}(A,B)=1-\frac{1}{m}\sum \limits_{i=1}^m{\left(\frac{1}{n_{x_{i}}}\sum\limits_{j=1}^{n_{x_{i}}}
{|h_{A}^{\sigma(j)}(x_{i})-h_{B}^{\sigma(j)}(x_{i})|^{p}}\right)}^{1/p},
\end{eqnarray}
\begin{eqnarray}\label{eqns5}
s_{3}(A,B)=1-\frac{1}{\sum \limits_{i=1}^m{(n_{x_{i}})^{1/p}}}\sum \limits_{i=1}^m{\left(\sum\limits_{j=1}^{n_{x_{i}}}
{|h_{A}^{\sigma(j)}(x_{i})-h_{B}^{\sigma(j)}(x_{i})|^{p}}\right)}^{1/p}.
\end{eqnarray}
where $p>0$.
\par If we take the weight of each element $x\in X$ into account, then we define the weighted similarity measures as:
\begin{eqnarray}\label{eqns43}
s_{4}(A,B)=1-\left[\sum \limits_{i=1}^m{w_{i}\left(\frac{1}{n_{x_{i}}}\sum\limits_{j=1}^{n_{x_{i}}}
{|h_{A}^{\sigma(j)}(x_{i})-h_{B}^{\sigma(j)}(x_{i})|^{p}}\right)}\right]^{1/p},
\end{eqnarray}
\begin{eqnarray}\label{eqns44}
s_{5}(A,B)=1-\sum \limits_{i=1}^m{w_{i}\left(\frac{1}{n_{x_{i}}}\sum\limits_{j=1}^{n_{x_{i}}}
{|h_{A}^{\sigma(j)}(x_{i})-h_{B}^{\sigma(j)}(x_{i})|^{p}}\right)}^{1/p},
\end{eqnarray}
\begin{eqnarray}\label{eqns6}
s_{6}(A,B)=1-\frac{m}{\sum \limits_{i=1}^m{(n_{x_{i}})^{1/p}}}\sum \limits_{i=1}^m{w_{i}\left(\sum\limits_{j=1}^{n_{x_{i}}}
{|h_{A}^{\sigma(j)}(x_{i})-h_{B}^{\sigma(j)}(x_{i})|^{p}}\right)}^{1/p}.
\end{eqnarray}
where $p>0$, $w_{i}(i=1,2,\cdots,m)$ with $w_{i}\in [0,1]$ and $\sum\limits_{i=1}^{m}{w_{i}}=1$.\\
\par Especially, if each element has the same importance, that is, $w_{i}=1/m, (i=1,2,\cdots,n)$, then the Eq.s (\ref{eqns43}), (\ref{eqns44}) and (\ref{eqns6}) are reduced to Eq.s (\ref{eqns3}), (\ref{eqns4}) and (\ref{eqns5}), respectively. \par
Let the universe of discourse $X=[a,b]$, the weight of element $x\in X$ be $w(x)$ with $w(x)\in [0,1]$ and $\int_{a}^{b}{w(x)dx}=1$, then we define the continuous similarity measures based on Eq.s (\ref{eqns43})-(\ref{eqns44}) as follow, respectively:
\begin{eqnarray}\label{eqns46}
s_{7}(A,B)=1-\int_{a}^{b}{w(x)\left(\frac{1}{n_{x}}\sum\limits_{j=1}^{n_{x}}
{|h_{A}^{\sigma(j)}(x)-h_{B}^{\sigma(j)}(x)|^{p}}\right)}^{1/p}dx
\end{eqnarray}
\begin{eqnarray}\label{eqns7}
s_{8}(A,B)=1-\frac{m}{\sum \limits_{i=1}^m{(n_{x_{i}})^{1/p}}}\int_{a}^{b}{w(x)\left(\sum\limits_{j=1}^{n_{x_{i}}}
{|h_{A}^{\sigma(j)}(x_{i})-h_{B}^{\sigma(j)}(x_{i})|^{p}}\right)}^{1/p}dx.
\end{eqnarray}
where $p>0$.\par
Specially, if $w(x)=1/(b-a)$ for all $x\in [a,b]$, then Eq.s (\ref{eqns46})-(\ref{eqns7})  becomes respectively\\
\begin{eqnarray}\label{eqns416}
s_{9}(A,B)=1-\frac{1}{b-a}\int_{a}^{b}{\left(\frac{1}{n_{x}}\sum\limits_{j=1}^{n_{x}}
{|h_{A}^{\sigma(j)}(x)-h_{B}^{\sigma(j)}(x)|^{p}}\right)}^{1/p}dx,
\end{eqnarray}
\begin{eqnarray}\label{eqns8}
s_{10}(A,B)=1-\frac{m}{(b-a)\sum \limits_{i=1}^m{(n_{x_{i}})^{1/p}}}\int_{a}^{b}{\left(\sum\limits_{j=1}^{n_{x_{i}}}
{|h_{A}^{\sigma(j)}(x_{i})-h_{B}^{\sigma(j)}(x_{i})|^{p}}\right)}^{1/p}dx.
\end{eqnarray}
where $p>0$.\par
It can be verified that $s_{i}(A,B)(i=4,6,\cdots,10)$ also have the properties (P1)-(P4).
\subsection{similarity measures based on the set-theoretic approach}
 The set-theoretic approach is used usually to similarity measures for fuzzy sets \cite{pappis1} and intuitionistic fuzzy sets \cite{xu3}. Thus we also define a similarity measure between two hesitant fuzzy sets $A$ and $B$ from the point of set-theoretic views as follows:
\begin{eqnarray}\label{eqns10}
s_{11}(A,B)=\frac{1}{m}\sum\limits_{i=1}^m{\frac{\sum\limits_{j=1}^{n_{x_{i}}}
{min\left(h_{A}^{\sigma(j)}(x_{i}),h_{B}^{\sigma(j)}(x_{i})\right)}}
{\sum\limits_{j=1}^{n_{x_{i}}}
{max\left(h_{A}^{\sigma(j)}(x_{i}),h_{B}^{\sigma(j)}(x_{i})\right)}}}
\end{eqnarray}
\begin{theorem}
$s_{11}(A,B)$ is a similarity measure of HFSs $A$ and $B$.
\end{theorem}
\noindent\textbf{Proof.}
It is obvious that $s_{11}(A,B)$ satisfies the properties (P1)-(P3). we only prove (P4). Let $A\sqsubseteq B\sqsubseteq C$, then $h_{A}(x_{i})\preceq h_{B}(x_{i})\preceq h_{C}(x_{i})$ for each $x_{i}\in X$. It follows that $0<h_{A}^{\sigma(j)}(x_{i})\leq h_{B}^{\sigma(j)}(x_{i})\leq h_{C}^{\sigma(j)}(x_{i})$ for all $x_{i}\in X$ and $j=1,2,\cdots,n_{x_{i}}$. Then we have
\begin{eqnarray*}
\begin{split}
&\frac{\sum\limits_{j=1}^{n_{x_{i}}}
{min\left(h_{A}^{\sigma(j)}(x_{i}),h_{C}^{\sigma(j)}(x_{i})\right)}}
{\sum\limits_{j=1}^{n_{x_{i}}}
{max\left(h_{A}^{\sigma(j)}(x_{i}),h_{C}^{\sigma(j)}(x_{i})\right)}}=\frac{\sum\limits_{j=1}^{n_{x_{i}}}
{h_{A}^{\sigma(j)}(x_{i})}}
{\sum\limits_{j=1}^{n_{x_{i}}}
{h_{C}^{\sigma(j)}(x_{i})}}\\
\leq &\frac{\sum\limits_{j=1}^{n_{x_{i}}}
{h_{A}^{\sigma(j)}(x_{i})}}
{\sum\limits_{j=1}^{n_{x_{i}}}
{h_{B}^{\sigma(j)}(x_{i})}}=
 \frac{\sum\limits_{j=1}^{n_{x_{i}}}
{min\left(h_{A}^{\sigma(j)}(x_{i}),h_{B}^{\sigma(j)}(x_{i})\right)}}
{\sum\limits_{j=1}^{n_{x_{i}}}
{max\left(h_{A}^{\sigma(j)}(x_{i}),h_{B}^{\sigma(j)}(x_{i})\right)}},\\
\Rightarrow &\frac{1}{m}\sum\limits_{i=1}^m{\frac{\sum\limits_{j=1}^{n_{x_{i}}}
{min\left(h_{A}^{\sigma(j)}(x_{i}),h_{C}^{\sigma(j)}(x_{i})\right)}}
{\sum\limits_{j=1}^{n_{x_{i}}}
{max\left(h_{A}^{\sigma(j)}(x_{i}),h_{C}^{\sigma(j)}(x_{i})\right)}}}\\
&\leq
\frac{1}{m}\sum\limits_{i=1}^m{\frac{\sum\limits_{j=1}^{n_{x_{i}}}
{min\left(h_{A}^{\sigma(j)}(x_{i}),h_{B}^{\sigma(j)}(x_{i})\right)}}
{\sum\limits_{j=1}^{n_{x_{i}}}
{max\left(h_{A}^{\sigma(j)}(x_{i}),h_{B}^{\sigma(j)}(x_{i})\right)}}}
\end{split}
\end{eqnarray*}
Thus, $s_{11}(A,C)\leq s_{11}(A,B)$. Similarly, we have $s_{11}(A,C)\leq s_{11}(B,C)$. $\Box$
\par If we take the weight of each element $x\in X$ into account, then we obtain
\begin{eqnarray}\label{eqns11}
s_{12}(A,B)=\sum\limits_{i=1}^m{w_{i}\frac{\sum\limits_{j=1}^{n_{x_{i}}}
{min\left(h_{A}^{\sigma(j)}(x_{i}),h_{B}^{\sigma(j)}(x_{i})\right)}}
{\sum\limits_{j=1}^{n_{x_{i}}}
{max\left(h_{A}^{\sigma(j)}(x_{i}),h_{B}^{\sigma(j)}(x_{i})\right)}}}
\end{eqnarray}
where $w_{i}\in [0,1]$ and $\sum\limits_{i=1}^{m}{w_{i}}=1$. Specially, if ${w_{i}}=1/m, (i=1,2,\cdots,m)$, then the Eq. (\ref{eqns11}) are reduced to the Eq. (\ref{eqns10}).\par
Let the weight of element $x\in X=[a,b]$ be $w(x)$ with $w(x)\in [0,1]$ and $\int_{a}^{b}{w(x)dx}=1$, then we define the continuous similarity measures corresponding to Eq. (\ref{eqns11}) as follow:
\begin{eqnarray}\label{eqns111}
s_{13}(A,B)=\int_{a}^{b}{w(x)\frac{\sum\limits_{j=1}^{n_{x}}
{min\left(h_{A}^{\sigma(j)}(x),h_{B}^{\sigma(j)}(x)\right)}}
{\sum\limits_{j=1}^{n_{x}}
{max\left(h_{A}^{\sigma(j)}(x),h_{B}^{\sigma(j)}(x)\right)}}}dx
\end{eqnarray}
Especially, if $w(x)=1/(b-a)$ for all $x\in [a,b]$, then Eq. (\ref{eqns111}) become\\
\begin{eqnarray}\label{eqns113}
s_{14}(A,B)=\frac{1}{b-a}\int_{a}^{b}{\frac{\sum\limits_{j=1}^{n_{x}}
{min\left(h_{A}^{\sigma(j)}(x),h_{B}^{\sigma(j)}(x)\right)}}
{\sum\limits_{j=1}^{n_{x}}
{max\left(h_{A}^{\sigma(j)}(x),h_{B}^{\sigma(j)}(x)\right)}}}dx
\end{eqnarray}
\par It is obvious that $s_{i}(A,B)(i=11,12,\cdots,14)$ also satisfies the properties (P1)-(P4).
\section{An application in multiple attribute decision making}
 In this section, we apply the above proposed similarity measures to multiple attribute decision making under hesitant fuzzy environment.
\par For a multiple attribute decision making problem, let $H=\{H_{1},h_{2},\cdots,h_{p}\}$ be a set of alternatives, $X=\{x_{1},x_{2},\cdots,x_{m}\}$ a set of attributes and $w=\{w_{1},w_{2},\cdots,w_{m}\}^{T}$ the weight vector of attributes, where
$w_{i}\in [0,1]$ and $\sum\limits_{i=1}^{m}{w_{i}}=1$.
\par Now we define respectively the notions of positive ideal $HFS$ and negative ideal $HFS$ as follows:
\begin{eqnarray}\label{eqnar90}
  H^{+}=\{ \frac{h_{H^{+}}(x_{i})}{x_{i}}|x_{i}\in X\}
\end{eqnarray}
and
\begin{eqnarray}\label{eqnar91}
  H^{-}=\{ \frac{h_{H^{-}}(x_{i})}{x_{i}}|x_{i}\in X\}
\end{eqnarray}
where
\begin{eqnarray*}
  h_{H^{+}}(x_{i})=\{h^{\sigma(k)}(x_{i})|h^{\sigma(k)}(x_{i})=\max_{j}\{h_{H_{j}}^{\sigma(k)}(x_{i})\},k=1,2,\cdots,n_{x_{i}}\},\\
  h_{H^{-}}(x_{i})=\{h^{\sigma(k)}(x_{i})|h^{\sigma(k)}(x_{i})=\min_{j}\{h_{H_{j}}^{\sigma(k)}(x_{i})\},k=1,2,\cdots,n_{x_{i}}\}.
\end{eqnarray*}
\par Based on the aforementioned formulae of similarity measures between $HFSs$,
we can calculate the degree of similarity of the positive ideal $HFS$ $H^{+}$ and alternative $H_{i}$, denoted by $s(H^{+},H_{i})$,
and the degree of similarity of the negative ideal $HFS$ $H^{+}$ and alternative $H_{i}$, denoted by $s(H^{-},H_{i})$, respectively.
\par Then we define the relative similarity measure $s_{i}$ corresponding to the alternative $H_{i}$ as follows:
\begin{eqnarray}\label{eqn188}
  s_{i}=\frac{s(H^{+},H_{i})}{s(H^{+},H_{i})+s(H^{-},H_{i})}, i=1,2,\cdots,m.
\end{eqnarray}
Obviously, the bigger the value $s_{i}$, the better the alternative $H_{i}$.

To illustrate the proposed similarity measures of $HFSs$ and the above approach of decision making, we give an example adapted from Example 1 in \cite{xu1} as follows:
\begin{example}
\end{example}
With the economic development of societies, energy is an essential factor. Therefore, the correct energy policy affects economic development and environment directly. Hence, the most appropriate energy policy selection is very important. Now we suppose that there are five energy projects as alternatives $H_{i}(i=1,2,3,4,5)$ to be invested, and four attributes $($$x_{1}$: technological; $x_{2}$: environmental; $x_{3}$: socio-political; $x_{4}$: economic$)$ to be considered. The weight vector of the attributes is $w=(0.15,0.3,0.2,0.35)^{T}$. Several decision makers are invited to evaluate the performances
of the five alternatives. For an alternative under an attribute, though all of the decision makers provide their evaluated
values, some of these values may be repeated. However, here we only consider all the possible values for an alternative under an attribute, that is to say these values repeated many times appear only once $($Xu and Xia explained the reason in \cite{xu1}$)$. In this case, all possible evaluations for an alternative under the attributes can be regarded as an $HFS$. For convenience, we use an hesitant fuzzy decision matrix to express the results evaluated by the decision makers, which is given in Table 1.
\begin{table}[H]\label{tab1}
\small
\begin{center}
\caption{Hesitant fuzzy decision making matrix}
{\begin{tabular}{@{}llllll}\hline
  \,  & {$x_{1}$} & {$x_{2}$} & {$x_{3}$} & {$x_{4}$}\\
\hline
 {$H_{1}$} & \{0.5,0.4,0.3\} & \{0.9,0.8,0.7,0.1\} & \{0.5,0.4,0.2\} & \{0.9,0.6,0.5,0.3\}\\
 {$H_{2}$} & \{0.5,0.3\} & \{0.9,0.7,0.6,0.5,0.2\} & \{0.8,0.6,0.5,0.1\} & \{0.7,0.3,0.4\}\\
 {$H_{3}$} & \{0.7,0.6\} & \{0.9,0.6\} & \{0.7,0.5,0.3\} & \{0.6,0.4\}\\
 {$H_{4}$} & \{0.8,0.7,0.4,0.3\} & \{0.7,0.4,0.2\} & \{0.8,0.1\} & \{0.9,0.8,0.6\}\\
 {$H_{5}$} & \{0.9,0.7,0.6,0.3,0.1\} & \{0.8,0.7,0.6,0.4\} & \{0.9,0.8,0.7\} & \{0.9,0.7,0.6,0.3\}\\
\hline
\end{tabular}}
\end{center}
\end{table}
If we use the formulae of similarity measure $s_{i}(A,B)(i=4,5,6,11)$ to calculate the degree of similarity between each alternative $H_{i}$ and the positive ideal alternative $H_{i}^{+}$ (or negative ideal alternative $H_{i}^{-}$), then we get
the rankings of these alternatives by Eq. (\ref{eqn188}). The results are listed in Tables 2-5, respectively.
We find that $H_{5}\succ H_{3}$ and they are superior to others whichever formula of similarity measure is used. From Tables 2-4, it is seen that, similar to literature \cite{xu1}, the rankings are different except Table 3 when the different values
of the parameter $p$ (which can be considered as the decision makers' risk attitude) are given. Therefore, the proposed similarity measures can provide the decision makers more choices according to the decision makers' risk attitudes and actual situations.
\begin{table}[H]\label{tab2}
\small
\begin{center}
\caption{Results obtained by the similarity measure $s_{4}(A,B)$.}
{\begin{tabular}{@{}lcccccccll}\hline
  \,  &\quad {$H_{1}$} &\quad {$H_{2}$} &\quad {$H_{3}$} &\quad {$H_{4}$}  &\quad {$H_{5}$} &\quad {Rankings}\\
\hline
 {$p=1$} &\quad 0.4719 &\quad 0.47033 &\quad 0.5111 &\quad 0.47788 &\quad 0.5547  &\quad $H_{5}\succ H_{3}\succ H_{4}\succ H_{1}\succ H_{2}$\\
 {$p=2$} &\quad 0.46814 &\quad 0.48052 &\quad 0.5138 &\quad 0.46197 &\quad 0.55475  &\quad $H_{5}\succ H_{3}\succ H_{2}\succ H_{1}\succ H_{4}$\\
 {$p=6$} &\quad 0.47238 &\quad 0.48158 &\quad 0.52557 &\quad 0.4262 &\quad 0.55783  &\quad $H_{5}\succ H_{3}\succ H_{2}\succ H_{1}\succ H_{4}$\\
 {$p=10$} &\quad 0.47854 &\quad 0.47206 &\quad 0.53101 &\quad 0.40649 &\quad 0.56777  &\quad $H_{5}\succ H_{3}\succ H_{1}\succ H_{2}\succ H_{4}$\\
\hline
\end{tabular}}
\end{center}
\end{table}

\begin{table}[H]\label{tab3}
\small
\begin{center}
\caption{Results obtained by the similarity measure $s_{5}(A,B)$.}
{\begin{tabular}{@{}lcccccccll}\hline
  \,  &\quad {$H_{1}$} &\quad {$H_{2}$} &\quad {$H_{3}$} &\quad {$H_{4}$}  &\quad {$H_{5}$} &\quad {Rankings}\\
\hline
 {$p=1$} &\quad 0.4719 &\quad 0.47033 &\quad 0.5111 &\quad 0.47788 &\quad 0.5547  &\quad $H_{5}\succ H_{3}\succ H_{4}\succ H_{1}\succ H_{2}$\\
 {$p=2$} &\quad 0.47016 &\quad 0.46967 &\quad 0.50993 &\quad 0.48055 &\quad 0.55334  &\quad $H_{5}\succ H_{3}\succ H_{4}\succ H_{1}\succ H_{2}$\\
 {$p=6$} &\quad 0.47058 &\quad 0.45747 &\quad 0.51003 &\quad 0.48376 &\quad 0.54219  &\quad $H_{5}\succ H_{3}\succ H_{4}\succ H_{1}\succ H_{2}$\\
 {$p=10$} &\quad 0.47124 &\quad 0.4518 &\quad 0.51049 &\quad 0.48481 &\quad 0.5389  &\quad $H_{5}\succ H_{3}\succ H_{4}\succ H_{1}\succ H_{2}$\\
\hline
\end{tabular}}
\end{center}
\end{table}

\begin{table}[H]\label{tab4}
\small
\begin{center}
\caption{Results obtained by the similarity measure $s_{6}(A,B)$.}
{\begin{tabular}{@{}lcccccccll}\hline
  \,  &\quad {$H_{1}$} &\quad {$H_{2}$} &\quad {$H_{3}$} &\quad {$H_{4}$}  &\quad {$H_{5}$} &\quad {Rankings}\\
\hline
 {$p=1$} &\quad 0.4728 &\quad 0.4735 &\quad 0.51883 &\quad 0.4735 &\quad 0.54951  &\quad $H_{5}\succ H_{3}\succ H_{2}\succ H_{4}\succ H_{1}$\\
 {$p=2$} &\quad 0.46962 &\quad 0.48329 &\quad 0.51937 &\quad 0.45865 &\quad 0.55016  &\quad $H_{5}\succ H_{3}\succ H_{2}\succ H_{1}\succ H_{4}$\\
 {$p=6$} &\quad 0.4976 &\quad 0.49856 &\quad 0.50208 &\quad 0.4905 &\quad 0.50783  &\quad $H_{5}\succ H_{3}\succ H_{2}\succ H_{1}\succ H_{4}$\\
 {$p=10$} &\quad 0.49985 &\quad 0.49978 &\quad 0.50015 &\quad 0.49819 &\quad 0.50167  &\quad $H_{5}\succ H_{3}\succ H_{1}\succ H_{2}\succ H_{4}$\\
\hline
\end{tabular}}
\end{center}
\end{table}

\begin{table}[H]\label{tab5}
\small
\begin{center}
\caption{Results obtained by the similarity measures based on the set-theoretic approach.}
{\begin{tabular}{@{}lcccccccll}\hline
  \,  &\quad {$H_{1}$} &\quad {$H_{2}$} &\quad {$H_{3}$} &\quad {$H_{4}$}  &\quad {$H_{5}$} &\quad {Rankings}\\
\hline
 {$s_{11}(A,B)$} &\quad 0.49857 &\quad 0.49975 &\quad 0.57059 &\quad 0.49975 &\quad 0.6122  &\quad $H_{5}\succ H_{3}\succ H_{2}\succ H_{4}\succ H_{1}$\\
\hline
\end{tabular}}
\end{center}
\end{table}
\section{Conclusion}
\hspace{6mm}In this paper, we have presented the modified axiom definitions of distance and similarity measure between
HFSs and proposed a series of hesitant distance measures based on
the Hamming distance, the Euclidean distance, $L_{P}$ metric and
exponential operation. We have also investigated the relationship between
distance measures and similarity measures, and according to
their relationships, the corresponding similarity measures between
HFSs have been obtained. Furthermore, we have also developed the similarity measures
 for HFSs based on set-theoretic approach and
 applied our similarity measures to hesitant fuzzy decision-making. The experiment results
 have showed that the proposed similarity measures and approach of decision making
 for HFSs are reasonable and efficient.

\section*{ Acknowledgments}

We would like to thank the anonymous reviewers for their helpful
comments and valuable suggestions.


\begin{thebibliography}{00}

\small
\bibitem{atanassov1} K. Atanassov, Intuitionistic fuzzy sets. Fuzzy Sets and Systems, 20: 87-96, 1986.
\bibitem{atanassov2} K. Atanassov, G. Gargov, Interval valued intuitionistic fuzzy sets, Fuzzy Sets and Systems, 31: 343-349, 1989.

\bibitem{candan1} K. Candan, W. Li, M. Priya, Similarity-based ranking and query processing in multimedia databases. Data \& Knowledge Engineering, 35: 259-298, 2000.
\bibitem{chen1} S. Chen, A new approach to handling fuzzy decisionmaking problems, IEEE Trans. Systems, Man, Cybernetics, 18: 1012-1016, 1988.
\bibitem{chen2}	N. Chen, Z. Xu, M. Xia, Correlation coefficients of hesitant fuzzy sets and their applications to clustering analysis, Applied Mathematical Modelling, 2012, in press, doi: 10.1016/j.apm.2012.04.031.
\bibitem{chen3}	S. Chen, Similarity measures between vague sets and between elements, IEEE Transactions on Systems, Man, and Cybernetics, Part B: Cybernetics, 27: 153-158, 1997.
\bibitem{chen4}	S. Chen, Measures of similarity between vague sets, Fuzzy Sets and Systems, 74: 217-223, 1995.

\bibitem{fan1} J.Fan, W. Xie, Some notes on similarity measure and proximity measure. Fuzzy Sets and Systems, 101: 403-412, 1999.
\bibitem{gau1} W. Gau, D. Buehrer, Vague sets. IEEE Trans. Systems,Man and Cybernetics. 23: 610-614, 1993.

\bibitem{hung1} W. Hung, M. Yang, Similarity measures of intuitionistic fuzzy sets based on Lp metric, International Journal of Approximate Reasoning, 46: 120-136, 2007.
\bibitem{hung2} W. Hung, M. Yang, Similarity measures of intuitionistic fuzzy sets based on Hausdorff distance, Pattern Recognition Letters, 25: 1603-1611, 2004.
\bibitem{hung3} W. Hung, M. Yang, On similarity measures between intuitionistic fuzzy sets, International Journal of intelligent systems, 23: 364-383, 2008.
\bibitem{hung4} W. Hung, Similarity measures between type-2 fuzzy sets, International Journal of Uncertainty, Fuzziness and Knowledge-Based Systems, 12: 827-841, 2004.

\bibitem{li1} D. Li, C. Cheng, New similarity measures of intuitionistic fuzzy sets and application to pattern recognitions, Pattern Recognition Letters, 23: 221-225, 2002.
\bibitem{li2} Y. Li, D. Olson, Z. Qin, Similarity measures between intuitionistic fuzzy (vague) sets: a comparative analysis, Pattern Recognition Letters, 28: 278-285, 2007.
\bibitem{liang1} Z. Liang, P. Shi, Similarity measures on intuitionistic fuzzy sets, Pattern Recognition Letters, 24: 2687-2693, 2003.
\bibitem{liu1} H. Liu, Multi-criteria deicision-making methods based on intuitionistic fuzzy sets, European Journal of Operational Research, 179: 220-233, 2007.
\bibitem{liu2} X. Liu, Entropy, distance measure and similarity measure of fuzzy sets and their relations, Fuzzy Sets and Systems, 52: 305-318, 1992.
\bibitem{mitchell1} H.B. Mitchell, On the dengfeng-chuntian similarity measure and its application to pattern recognition, Pattern Recognition Letters 24, 3101-3104, 2003.
\bibitem{pal1} N. Pal, S. Pal, Some properties of the exponential
entropy. Information Science, 66: 119-137, 1992.
\bibitem{pappis1} C. Pappis, N. Karacapilidis, A comparative assessment of measures of similarity of fuzzy values, Fuzzy Sets and Systems, 56: 171-174, 1993.

\bibitem{rodriguez1} R. Rodriguez, L. Martinez, F. Herrera, Hesitant fuzzy linguistic term sets for decision making, IEEE Transactions on Fuzzy Systems, 20: 109-119, 2012.

\bibitem{torra1}	V. Torra, Hesitant fuzzy sets, International Journal of Intelligent Systems, 25: 529-539, 2010.
\bibitem{torra2}	V. Torra, Y. Narukawa, On hesitant fuzzy sets and decision, in: The 18th IEEE International Conference on Fuzzy Systems, Jeju Island, Korea, 2009, 1378-1382.
\bibitem{turksen1} I. Turksen, Z. Zhong, An approximate analogical reasoning approach based on similarity measures, IEEE Transactions on Systems, Man and Cybernetics 18: 1049-1056, 1988.

\bibitem{wang1} W. Wang, New similarity measures on fuzzy sets and on elements, Fuzzy Sets and Systems, 85: 305-309, 1997.


\bibitem{wei1}	G. Wei, Hesitant fuzzy prioritized operators and their application to multiple attribute decision making , Knowledge-Based Systems, 31: 176-182, 2012.

\bibitem{xia1}	M. Xia and Z. Xu, Hesitant fuzzy information aggregation in decision making, International Journal of Approximate Reasoning, 52, 395-407, 2011.
\bibitem{xia2} M. Xia, Z. Xu, N. Chen, Some hesitant fuzzy aggregation operators with their application in group decision making, Group Decision and Negotiation, 2012, in press, doi: 10.1007/s10726-011-9261-7.

\bibitem{xu1} Z. Xu, M. Xia, Distance and similarity measures for hesitant fuzzy sets, Information Sciences, 181: 2128-2138, 2011.
\bibitem{xu2}	Z. Xu, M. Xia, On distance and correlation measures of hesitant fuzzy information, International Journal of Intelligent Systems, 26, 410-425, 2011.
\bibitem{xu3} Z. Xu, Some similarity measures of intuitionistic fuzzy sets and their applications to multiple attribute decision making, Fuzzy Optimization and Decision Making, 6: 109-121, 2007.
\bibitem{xu4} Z. Xu, J. Chen, Ordered weighted distance measure. Journal of Systems Science and Systems Engineering, 16: 529-555, 2008.
\bibitem{xu5}	Z. Xu, M. Xia, Hesitant fuzzy entropy and cross-Entropy and their use in multiattribute decision-making, International Journal of Intelligent system, 27: 799-822, 2012.


\bibitem{yang1} M. Yang, K. Wu, A similarity-based robust
clustering method. IEEE Trans. Pattern Anal. Machine
Intell. 26: 434-448, 2004.
\bibitem{yang2} M. Yang, D. Lin, On similarity and inclusion measures between type-2 fuzzy sets with an application to clustering. Computers \& Mathematics with Applications, 57: 896-907, 2009.
\bibitem{yu1}	D. Yu, Y. Wu, W. Zhou, Multi-criteria decision Making based on Choquet integral under hesitant fuzzy environment, Journal of Computational Information Systems, 7: 4506-4513, 2011.

\bibitem{zadeh1} L. Zadeh, Fuzzy Sets. Information and Control, 8: 338-353, 1965.

\bibitem{zadeh2} L. Zadeh, The concept of a linguistic variable and its application to approximate reasoning-I. Information Sciences 8: 199-249, 1975.
\bibitem{zadeh3} L. Zadeh, Similarity relations and fuzzy orderings.
Information Science, 3: 177-200, 1971.
\bibitem{zeng1} W. Zeng, H. Li, Relationship between similarity and entropy of interval
valued fuzzy sets, Fuzzy Sets and Systems, 157: 1477-1484, 2006.
\bibitem{zhang1} C. Zhang, H. Fu, Similarity measures on three kinds of fuzzy sets, Pattern Recognition Letters, 27: 1307-1317, 2006.
\bibitem{zhu1} B. Zhu, Z. Xu, M. Xia, Hesitant fuzzy geometric Bonferroni means, Information Sciences, 2012, in press, doi: 10.1016/j.ins.2012.01.048.

\end{thebibliography}
\end{document}